\documentclass[
    ,final            
  ]
  {aipproc}

\layoutstyle{6x9}

\begin{document}

\title{Flaring variability of Microquasars}

\classification{95.85.Bh; 95.85.Nv; 97.80.Jp; 97.60.Lf}

\keywords {X-ray binaries, microquasars,radio astronomy, X-rays astronomy}

\author{Sergei A. Trushkin, Nikolaj N. Bursov, Nikolaj A. Nizhelskij}{
  address={Special Astrophysical Observatory RAS, Nizhnij Arkhyz, 369167, Russia}
}

\begin{abstract}
We discuss flaring variability of radio emission of microquasars,
measured in monitoring programs with the RATAN-600 radio telescope.
We carried out a multi-frequency (1-30 GHz) daily monitoring of the
radio flux variability of the microquasars SS433, GRS1915+105, and
Cyg X-3 during the recent sets in 2005-2007.
A lot of bright short-time flares were detected from GRS~1915+105 and they
could be associated with active X-ray events.
In January 2006 we detected a drop down of the quiescent fluxes from Cyg X-3
(from 100 to $\sim$20 mJy), then the 1~Jy-flare was detected on 2 February
2006 after 18 days of quenched radio emission. The daily spectra
of the flare in the maximum were flat from 2 to 110 GHz, using the
quasi-simultaneous observations at 110 GHz with the RT45m telescope and
the NMA millimeter array of NRO in Japan.
Several bright radio flaring events (1-15 Jy) followed during the continuing
state of very variable and intensive 1-12 keV X-ray emission ($\sim$0.5 Crab),
which was monitored in the RXTE ASM program. Swift/BAT ASM hard X-ray fluxes
correlated strongly with flaring radio data. The various
spectral and temporal characteristics of the light curves from the
microquasars could be determined from such comparison.
We conclude that monitoring of the flaring radio emission is a good tracer
of jet activity X-ray binaries.
\end{abstract}

\maketitle

\subsection{Introduction}

The first idea about cosmic radio sources variability belongs
to Shklovskij, who predicted the secular decreasing of the total radio
flux from the SNR Cas A \cite{shk60}.
Strong variability of the cosmic sources was detected from extragalactic
sources, quasars and AGNs in 60th years, and later from SS433,
prototype radio emitted X-ray binary with merely relativistic jets.
Van der Laan developed Shklovskij's idea to generalize the main
formulae and showed that any synchrotron emitting source should evolve in
a similar manner \cite{laan66}. Marscher and Gear \cite{mg85} were the
first who to use Rees's idea \cite{rees78} about internal shocks, running
in the jets of flaring quasar 3C273.

Variable synchrotron emission in microquasars,
quasars and AGNs is originated from outflows of accreting matter in the
narrow cone -- the two-side relativistic jets, ejected from polar regions
of accretion disks around black holes or neutron stars.
The jets contain magnetic fields and energetic electrons.
The temporal and frequency dependencies in the light curves are a key
for clear understanding and good probe test for models of the physical
processes in cosmic jets. A comparison of the radio and X-ray variable
emission allows us to provide detailed studies.

\subsection{Observations}

We have carried out the almost daily monitoring observational sets
of the microquasars Cyg X-3, GRS~1915+10, SS433
from September 2005 to May 2006, in July 2006, from November 2006
to March 2007 and from December 2007 to February 2008 at frequencies
from 1 to 30 GHz.
The measured multi-frequency light curves can be directly compared with time
series of the X-ray observatory RXTE \cite{lev96} and hard X-ray data from
Swift/BAT (15-50 keV)\cite{bat05}.
The observations have been made with the `Northern sector' antenna of
the RATAN-600 radio telescope in a transit mode.
As a rule the errors of measured fluxes of the sources did not exceed
5-10 per cent. The details of the observations, the errors are given in
\cite{tru06a}.

\subsection{GRS 1915+105: X-ray -- radio correlation}

The X-ray transient source GRS~1915+105 was discovered in 1992 by
Castro-Tirado et al. \cite{CT92} with the WATCH instrument on board GRANAT.
An apparent super-luminous motion of the jet components was detected
and  the determination of 'microquasars' were coined \cite{mr94}.
Fender et al. \cite{fender02} discussed the alert observations of two
flares (July 2000), when for the first time detected the quasi-periodical
oscillations with P = 30.87 minutes at two frequencies: 4.8 and 8.64 GHz.
Linear polarization was measured at a level 1-2 per cent
at both frequencies as well.

\begin{figure}
\includegraphics[width=0.51\linewidth]{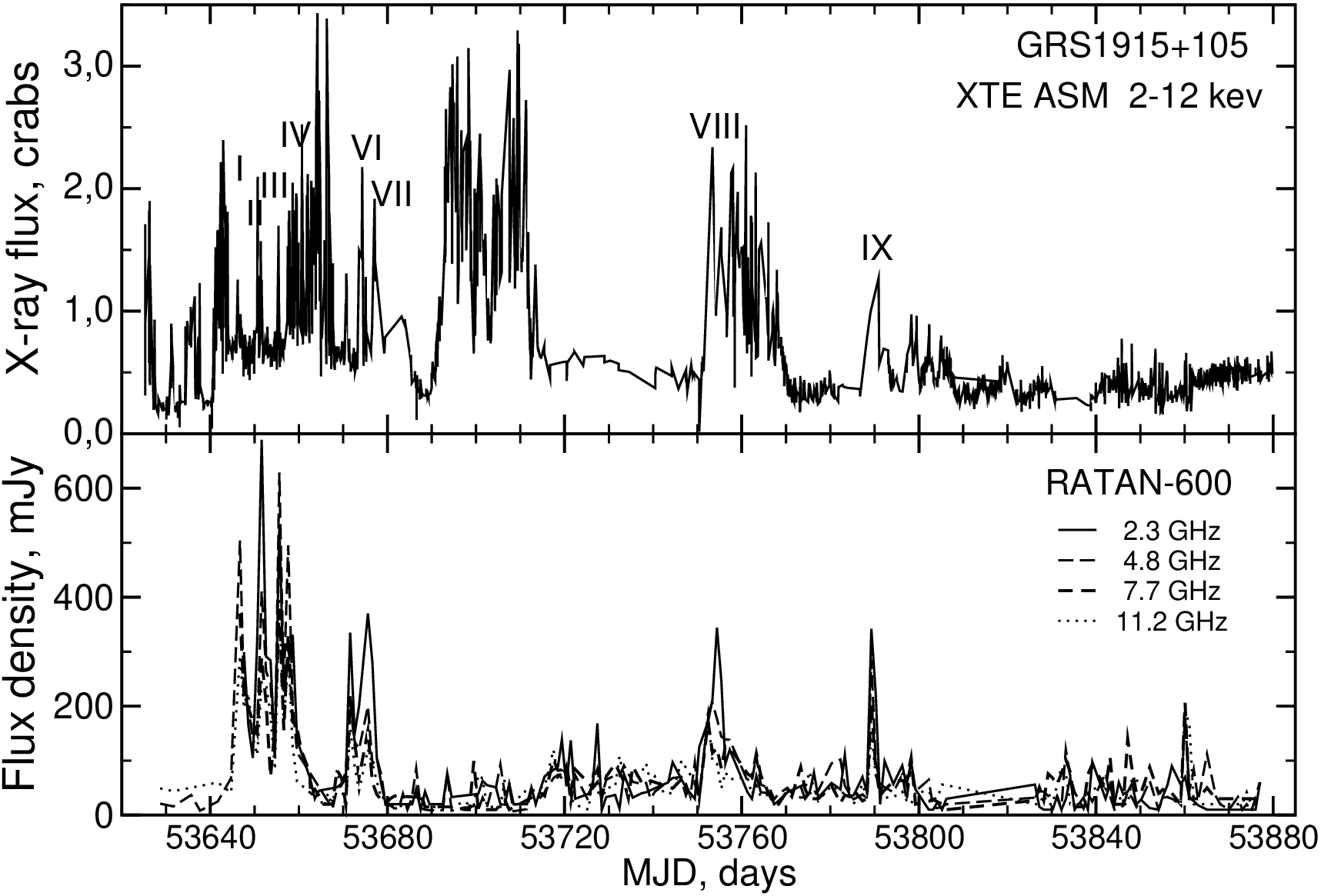}
\includegraphics[width=0.48\linewidth]{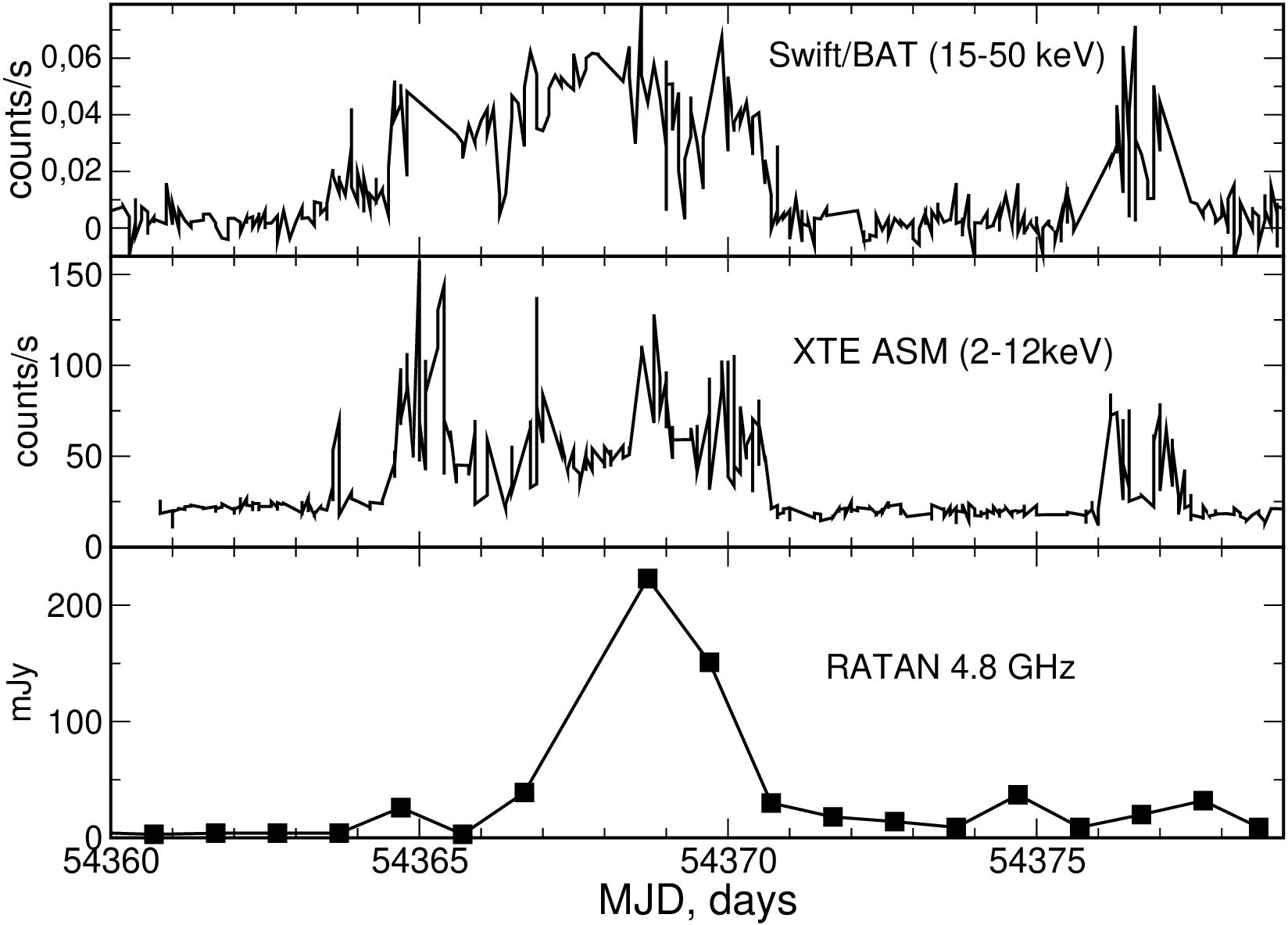}
\caption{Light curves of GRS1915+105 at radio frequencies and at 2-12 keV
from September 2005 to March 2006 and
light curves of GRS1915+105 at radio frequencies and X-ray  2-12 keV and 15-50
keV fluxes in September 2007.}
\label{fig:1}
\end{figure}

In Fig.\ref{fig:1}{(\it left)} the radio and X-ray light curves are showed
for the total set of 2005-2006. The nine radio flares have the counterparts
in X-rays. The radio spectrum was optically thin in the first two flares,
and optically thick in the third one (see details in \cite{nam07}).
In Fig.\ref{fig:1} {(\it right)}
the radio and sort and hard X-ray light curves during the September-October
2007. Again we see detectable correlation between the radio flare and
a soft X-ray 'spike' in the high state.

The profiles of the X-ray spikes during the radio flares
are clearly distinguishable from other spikes because its shape
shows a fast-rise and a exponential-decay. The X-ray spikes,
which reflect activity of the accretion disk, show an irregular
pattern. During a bright radio flare, the spectra of the X-ray spikes become
softer than those of the quiescent phase, by a fraction of $\sim$30\%
\cite{nam07}. But such rule is not universal.

\begin{figure}
\centering
\includegraphics[width=0.5\linewidth]{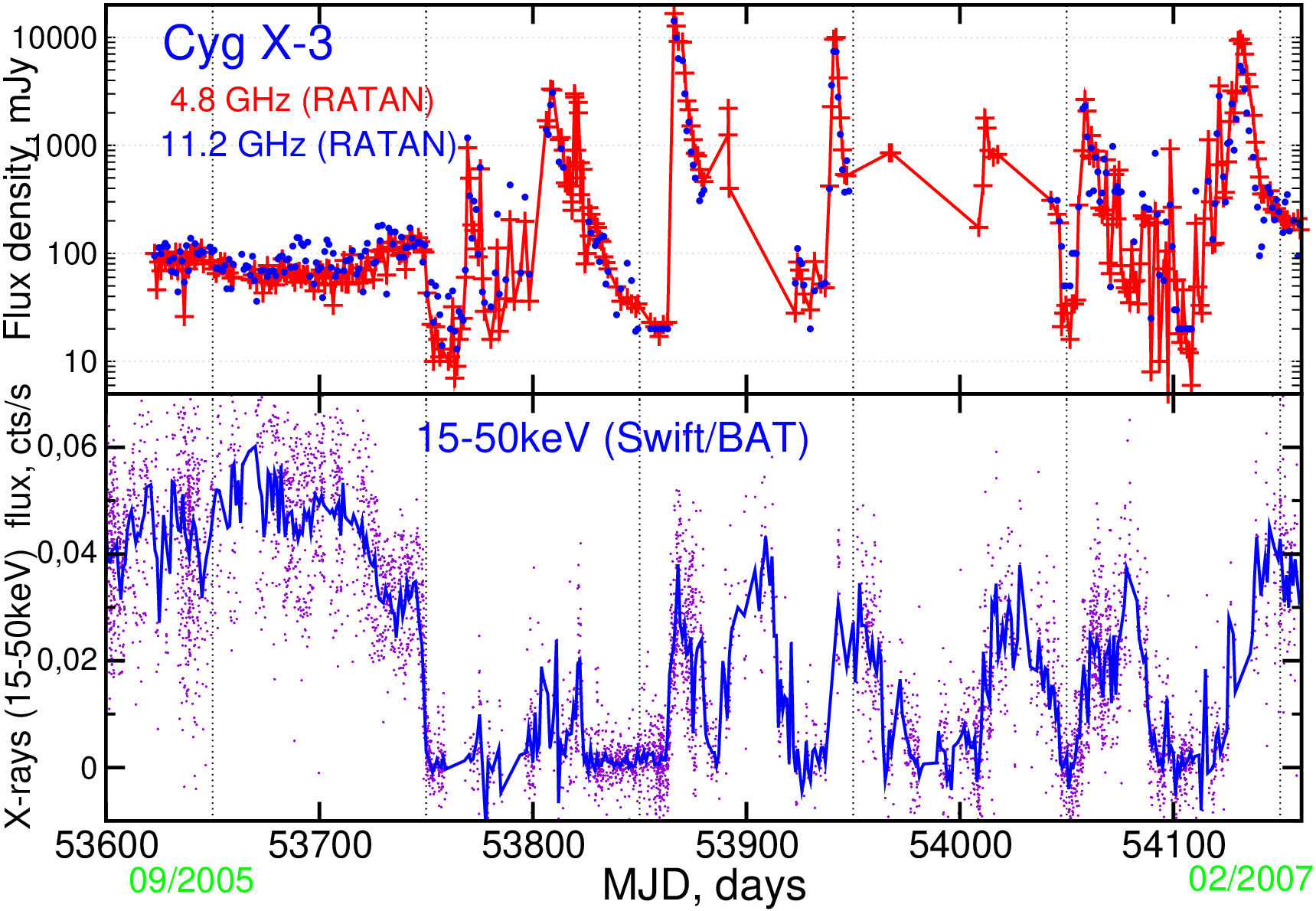}
\includegraphics[width=0.5\linewidth]{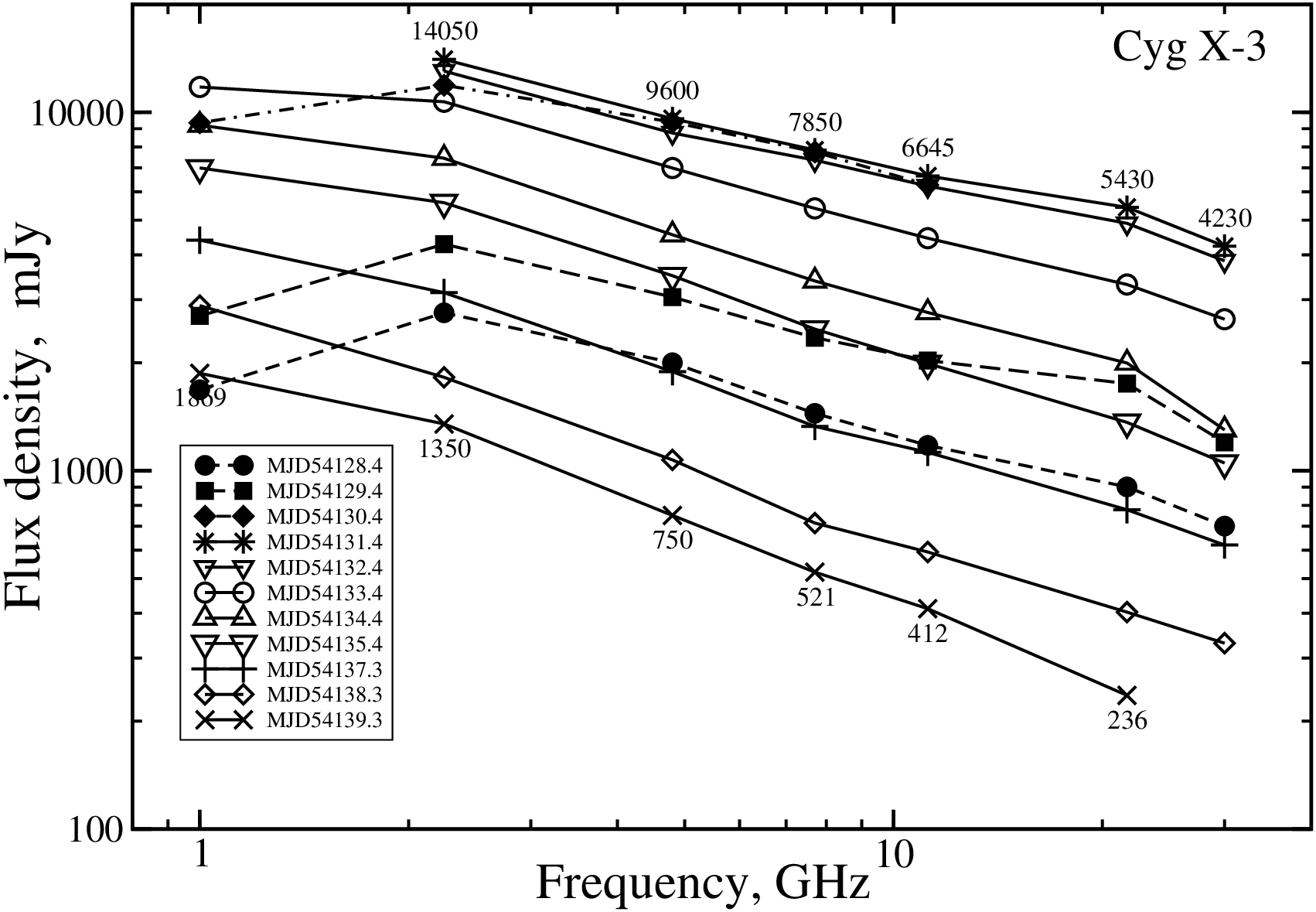}
\caption{{\it (right)} The RATAN and Swift/BAT light curves of Cyg X-3 from September 2005 to
February 2007 and {(\it left)}: The daily spectra of Cyg X-3 during the flare in Jan-Feb 2007.}
\label{fig:2}
\end{figure}

Miller at al. \cite{miller06} have detected a one-side large-scale radio jet
from GRS1915+105 with VLBA mapping during an X-ray and (IX) radio outburst
on 23 February 2006 (MJD53789.258). Then the optically thin flare with fluxes
340, 340, 342, 285, 206, and 153 mJy was detected at frequencies
1, 2.3, 4.8, 7.7, 11,2 and 21.7 GHz respectively.

\subsection{Cyg X-3: 2006-2007 -- a new long-term active period }

During ~100 days (September -- December 2005)
Cyg~X-3  was in a quiescent state of $\sim$100 mJy (Fig.\ref{fig:2}).
In December 2005 its X-ray flux began to increase and the radio flux at 2-11
GHz increased also. Then the flux density of the source at 4.8 GHz  was
found to drop from 103 mJy on 2006 Jan 14.4 (UT) to 43 mJy on Jan 15.4 (UT),
and to 22 mJy on Jan 17.4 (UT). The source is known to exhibit radio
flares typically with a few peaks exceeding 1-5 Jy following such a quenched
state as Waltman et al. \cite{waltman94} have showed in
the intensive monitoring of Cyg X-3 with the Green Bank
Interferometer at 2.25 and 8.3 GHz.
The source has been monitored from 2006 Jan 25 (UT) with the Nobeyama Radio
Observatory 45m Telescope (NRO45m Telescope),
the Nobeyama Millimeter Array (NMA) and Japanese VLBI Network telescopes.
On Feb 2.2 (UT), about 18 days after it entered the quenched state, the
rise of a first peak is detected with the NRO45m Telescope and YRT32m.
On Feb 3.2 (UT), the flux densities reached the first peak at all the
sampling frequencies from 2.25 GHz to 110.10 GHz (\cite{tsuboi06}).
The spectrum at maximum (3 February) of the flare was flat
as measured by RATAN, NRO RT45m and NMA from 2 to 110 GHz.
The next peak of the active events on 10 February
reached the level of near 1 Jy again with a similar flat spectrum.
Then three short-time flares have happened during a week.
The flare on 18 February had the inverted spectrum with the same
spectral index $\alpha$=+0.75 from 2.3 to 22 GHz.

\begin{figure}
\centering
\includegraphics[width=0.5\linewidth]{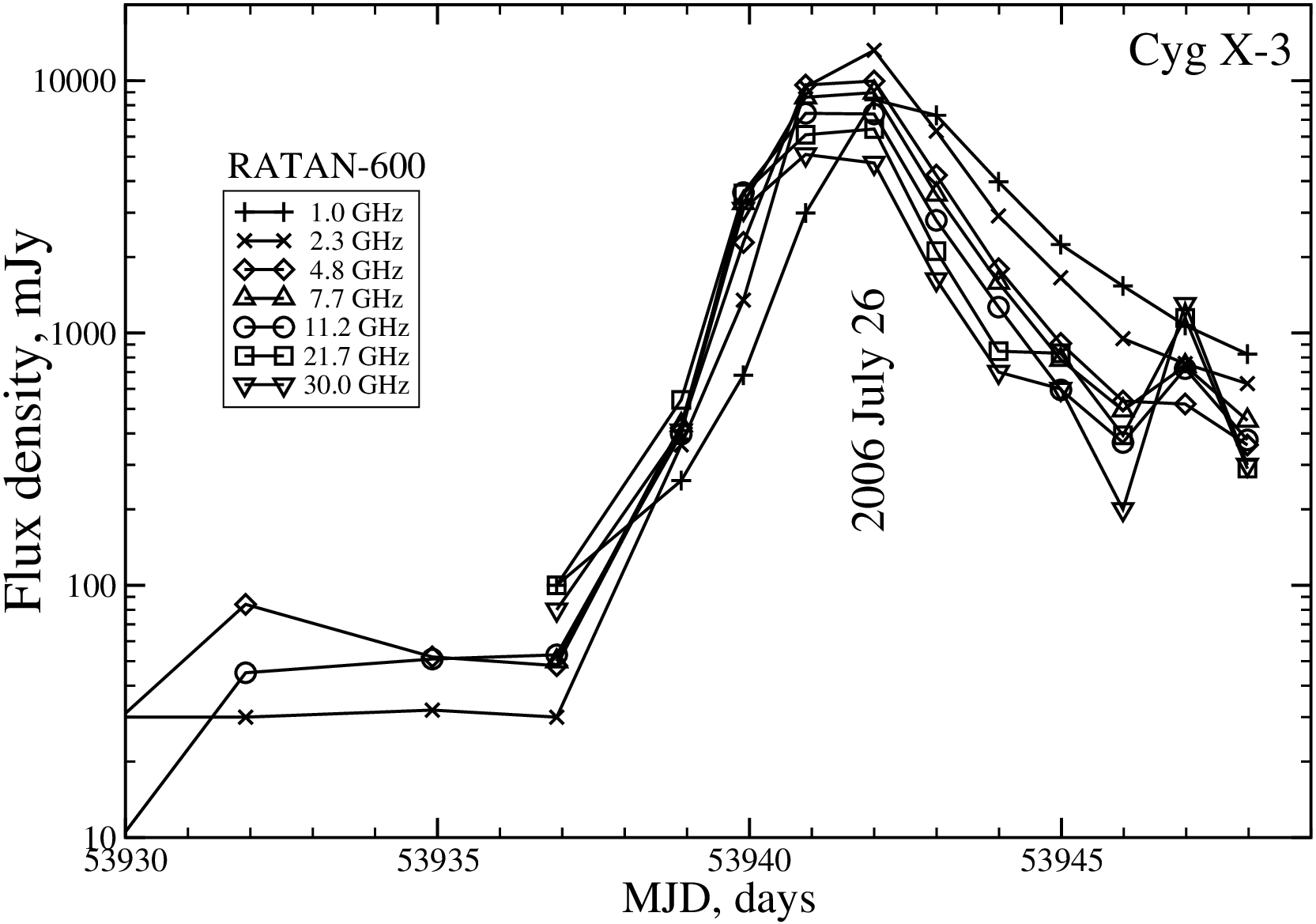}
\includegraphics[width=0.5\linewidth]{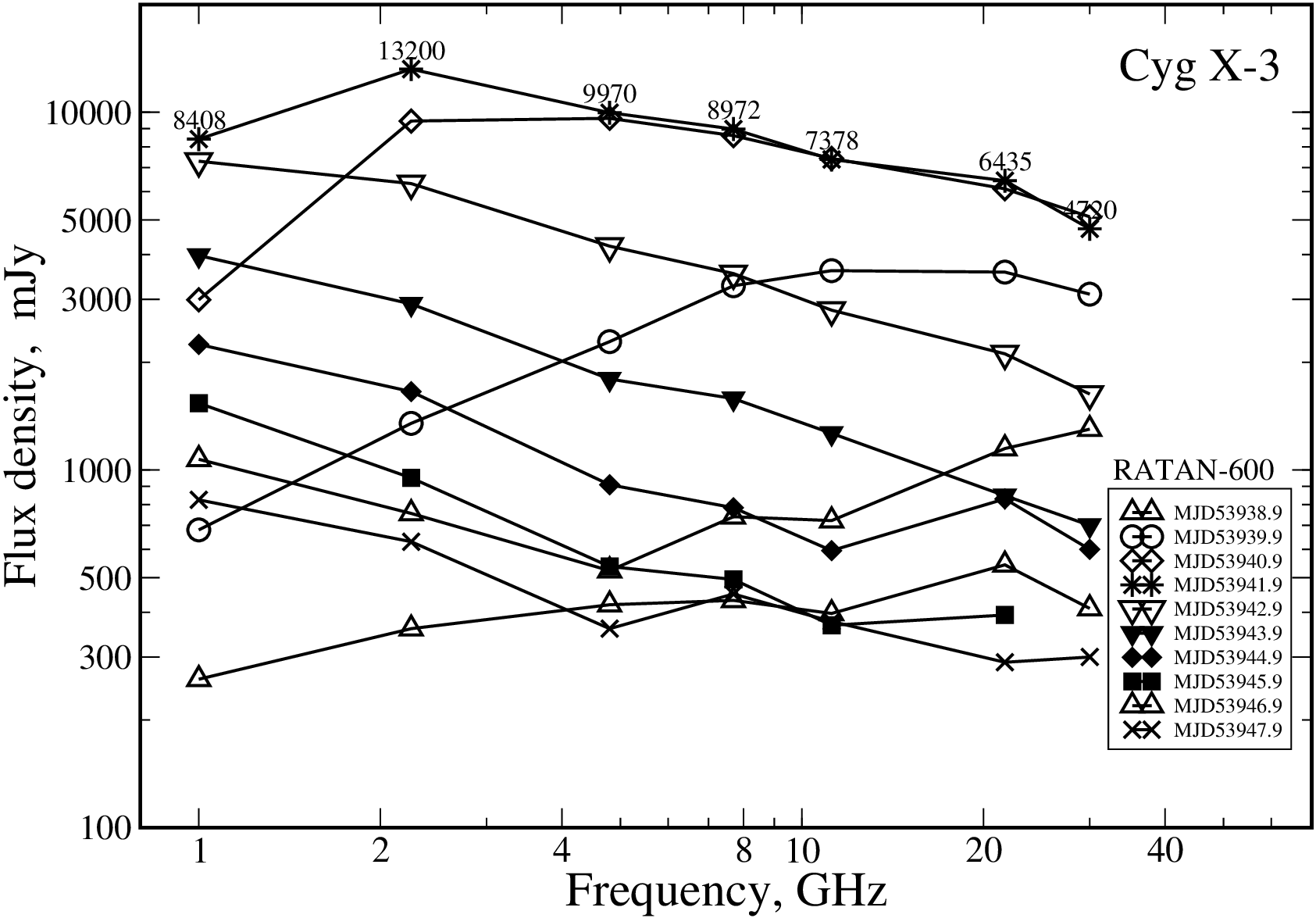}
\caption{The RATAN and RXTE ASM light curves of Cyg X-3 in July 2006 (right) and
the daily spectra of Cyg X-3 during flare in July 2006 (left).}
\label{fig:3}
\end{figure}

In the active period (2006) there were two powerful flares,
March 14 to 3-5 Jy and May 11 to 12-16 Jy at 2-30 GHz. In the May flare
fluxes have grown up by a factor $\sim$1000 during a day.
The change of the spectrum  during the flare on May 11-19
followed the model of a single ejection of the relativistic electrons, moving
in thermal matter in the intensive WR-star wind.
It stays in optically thin mode at  higher frequencies,
meanwhile at lower frequency 614 MHz \cite{pal06} Cyg X-3 was in hard absorption.
In the continuing active state of Cyg X-3 we detected a very fast-rise flare
at 2.3 and 8.5 GHz with RT32 (IAA) on 5 June 2006 (MJD53891) \cite{tru06b}.
During 3 hours the fluxes changed from $\sim1$ Jy to 2 Jy and then
decreased to 100-400 mJy during 15 hours.
We detected the similar behaviour in the the Jan-Feb 2007, when maximum
flux reached 14 Jy at 2.3 GHz (Fig:2).

In Fig.\ref{fig:3} the light curves of the July flare are shown.
The phase of the flux rising continued during 4-5 days.
For the first time we could clearly see the evolution of the spectrum  during
the phase of the flare (Fig.\ref{fig:3}). And it was amazing that
the low-frequency part of the spectra evolved from nearly flat optically
thin (at 1 GHz) on the first day to optically thick after 3-4 days.
In the standard model of the expansion of the compact
sources (jets components) there is no any explanation for such behaviour.
We had to conclude that the thermal electron density, responsible for the
low frequency absorption, grows up during grow-up of the relativistic electron
density.
The later stage of the spectral and temporal evolution could be
fitted by the modified finite segments model by Marti et al. \cite{marti92} or
Hjellming \cite{Hj88} and Hjellming et al. \cite{Hj00}.
Indeed in Fig.\ref{fig:3} the radio spectra of the July flare
evolved from the fourth day (MJD53942) as usual adiabatically expanding
relativistic jets moving with $v_{jet}\sim0.74c$,
and thermal electron component has: $T_e = 10^4$ K, $n_{th} = 2\,\,10^4 cm^{-3}$,
magnetic field $B_0 = 0.07$ Gs, and energy index $\gamma = -1.85$.
During the rising stage of the flare we should involve
the intense internal shocks running through the jet \cite{wad03}.

\begin{figure}
\includegraphics[width=0.5\linewidth]{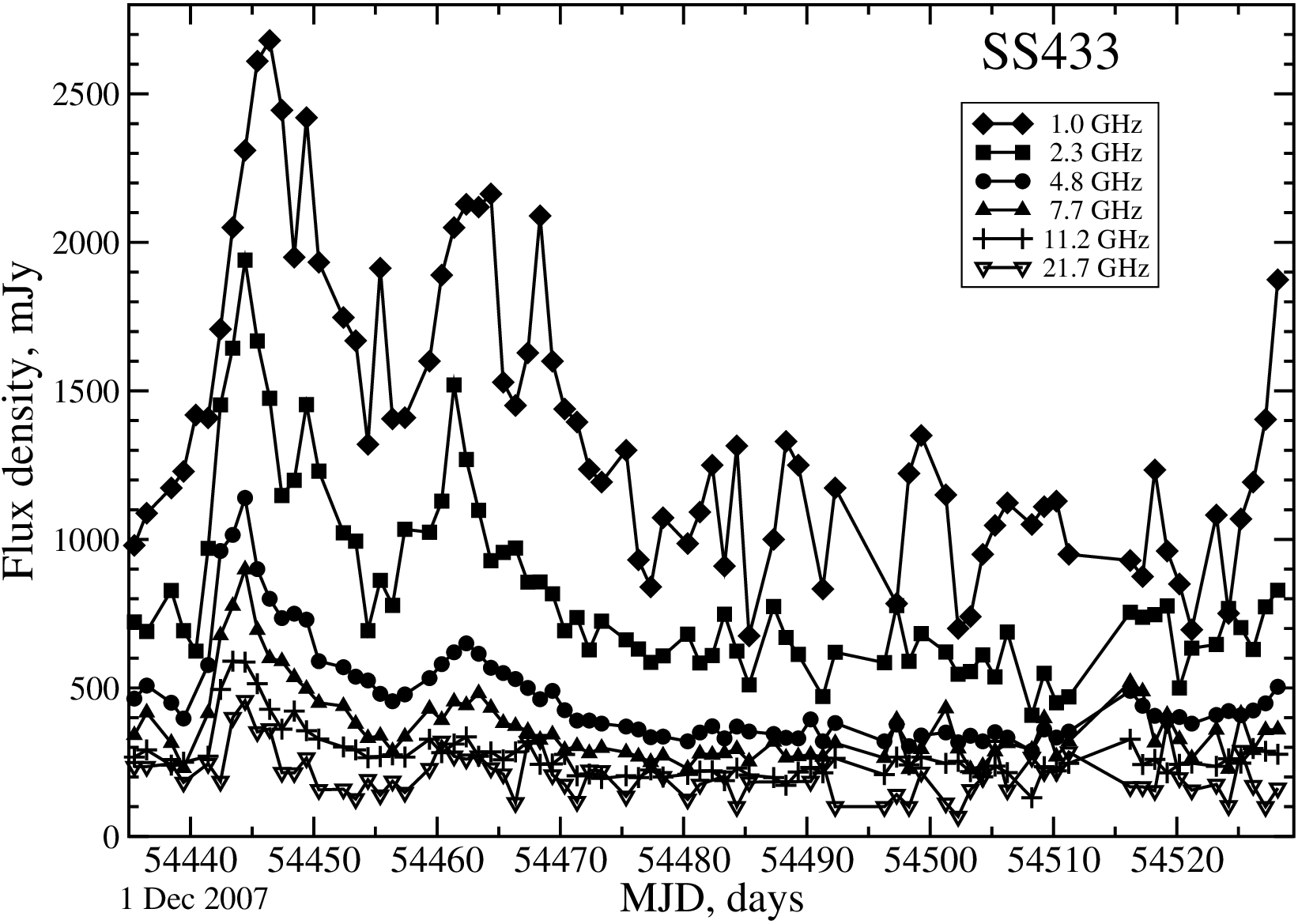}
\includegraphics[width=0.5\linewidth]{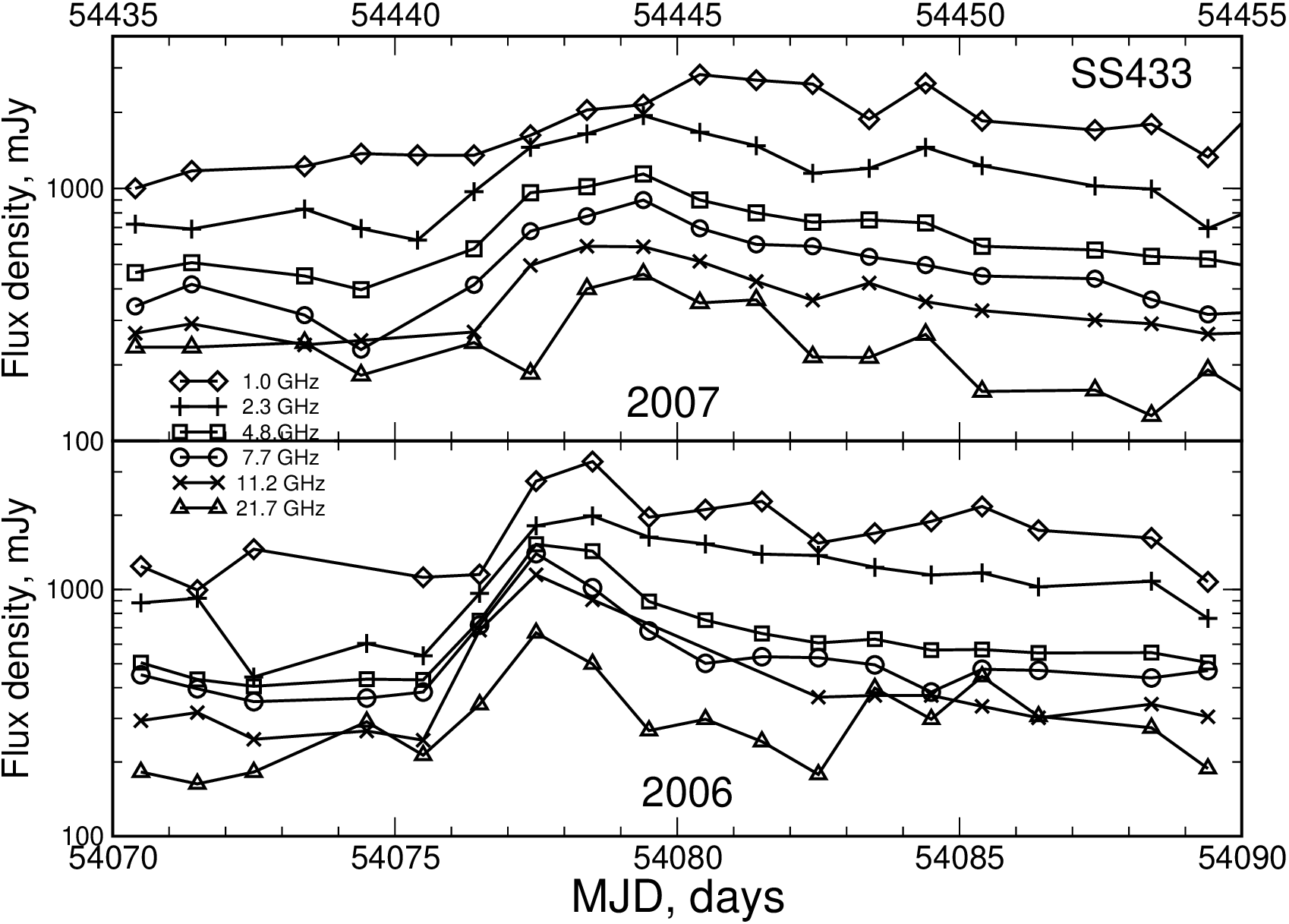}
\caption{The light curves of SS433
in December 2007 - January 2008 during the powerful flaring events.
and comparison two flare in December 2006 and December 2007}

\label{fig:4}
\end{figure}

\subsubsection{SS433: new flaring events and their spectra}

The first microquasar SS433, a bright variable emission star was
identified  with a rather bright compact radio
source 1909+048 located in the center of a supernova remnant W50.
In 1979 moving optical emission lines, Doppler-shifted due to precessing
mass outflows with 78000 km/s, were discovered in the spectrum of
SS433. At the same time in 1979 were discovered a unresolved
compact core and 1 arcsec long aligned jets in the MERLIN radio
image of SS433. Since 1979  many monitoring sets
(for ex., with GBI \cite{fiedler87b}, RATAN \cite{tru03}) were began.
Different data do indicate a presence of a very
narrow (about 1$^o$) collimated beam at least in X-ray and optical
ranges. At present there is no doubt that SS433 is related to W50.
A distance of near 5 kpc was later determined
by different ways including the direct measurement of proper motions of
the jet radio components.

Kotani et al. \cite{kotani06} detected the fast variation in the X-ray
emission of SS433 during the radio flares, and probably QPOs of 0.11 Hz.
Massive ejections during this active period could  be the reason of
such behavior.
The daily RATAN light curves are measured from September 2005 to May 2006.
The activity of SS433 began during the second half of the monitoring
set. Some flares happened just before and after the multi-band program
of the studies of SS433 in April 2006 \cite{kotani07}.

In Fig.\ref{fig:4}(left) the light curves during the bright flare in
December 2007 are showed.
The delay of the maximum flux at 1 GHz is about 2 days and 1 day at 2.3  GHz
relative to the maxima at higher frequencies.

In Fig.\ref{fig:4}(right) the light curves during the bright flares in
December 2006 and 2007 are showed. The surprising coincident dates (6-7 Dec)
of both flares could be evidence of the 1-year periodicity of activity
in SS433. Nandi et al. \cite{nandi05} discussed the periodicity of flaring
events in ASM RXTE X-ray data, and found possible period about 368 days.

\subsection{Conclusions}

The RATAN microquasar monitoring data give us abundant material for
comparison with X-ray data from the ASM or ToO programs with RXTE,
CHANDRA, Suzaku and INTEGRAL. The 1-30 GHz emission originates often
from different optically thin and thick regions and we proposed an adequate
model of the flaring radio emission producing in the relativistic jets
interacting with varying circumstellar medium or intense stellar wind.

\begin{theacknowledgments}
We acknowledge the use of public data from the Swift data archive.
These studies are supported by the Russian  Foundation  Base Research (RFBR)
grant N~08-02-00504 and the mutual RFBR and
Japan Society for the Promotion of Science (JSPS) grant N~05-02-19710.
\end{theacknowledgments}

\end{document}